# Structural investigation of the liquid crystalline phases of three homologues from the nOS5 series (n = 9, 10, 11) by X-ray diffraction

Aleksandra Deptuch[1,*], Bartosz Sęk[2], Sebastian Lalik[3], Mirosława D. Ossowska-Chruściel[4], Janusz Chruściel[4], Monika Marzec[3]

[1] Institute of Nuclear Physics Polish Academy of Sciences, Radzikowskiego 152, PL-31342 Kraków, Poland
[2] Faculty of Physics and Applied Computer Science, AGH University of Kraków, Reymonta 19, PL-30059 Kraków, Poland
[3] Institute of Physics, Jagiellonian University, Łojasiewicza 11, PL-30348 Kraków, Poland
[4] Institute of Chemistry, Siedlce University of Natural Sciences and Humanities, 3 Maja 54, PL-08110 Siedlce, Poland
* corresponding author, aleksandra.deptuch@ifj.edu.pl

## Abstract

Polarizing optical microscopy and differential scanning calorimetry are used to determine the phase sequence of three liquid crystalline 4-pentylphenyl-4'-n-alkyloxythiobenzoates with n = 9, 10, 11. The X-ray diffraction method is applied for structural characterization of the liquid crystalline phases. The smectic layer spacing, tilt angle, average distance between the long axes of molecules and correlation length of the short-range order are determined as a function of temperature. For the crystal-like smectic phases with the hexagonal or herring-bone packing, the unit cell parameters are obtained. The presence of the tilted hexagonal phase for n = 10, 11 and tilted herring-bone phase for n = 9, 10 is indicated, although the direction of the tilt cannot be determined.

**Keywords:** thermotropic liquid crystals, hexagonal smectic phases, herring-bone smectic phases, layer spacing, X-ray diffraction

## 1. Introduction

The main thermotropic liquid crystalline phases created by rod-like molecules are the nematic phase and several smectic phases. The nematic phase, denoted as N, shows the long-range orientational order and liquid-like (short-range) positional order. In the smectic phases, denoted as SmX (where X stands for A, C, $B_{hex}$, F, I, B, G, J, E, H, K), there is the positional lamellar order [1-3]. In the orthogonal smectics A, $B_{hex}$, B, E, the average tilt angle of long molecular axes in respect to the layer normal is equal to zero, while in the tilted smectics C, F, I, G, J, H, K, the average tilt angle is larger than zero. Another difference between the smectic phases is the order within layers. In smectics A and C, it is only a liquid-like order. In the hexatic smectics $B_{hex}$, F and I, the positional order is also liquid-like (with a larger correlation length than in SmA and C), however, the bond-orientational order arises, which means that clusters with a short-range hexagonal order are oriented in the same direction. In smectics B, G, J, there is a long-range hexagonal order within smectic layers, while in smectics E, H, K it is a long-range herring-bone order [1,2]. The structures of smectics with the hexagonal or herring-bone order can be described by unit cells, thus they are referred to also as soft crystals [1,2]. In this work, we prefer to keep the "Sm" notation for the B, G, J, E, H, K phases, to distinguish them from the crystal phases with the orientational order.

In the homologous series of liquid crystalline compounds, the length of the $C_nH_{2n}$ or $C_nH_{2n+1}$ chain usually has a significant influence on the mesomorphic properties (presence of certain phases and phase transition temperatures), to give as example papers [4-10]. The same is in the case of the homologous series of 4-pentylphenyl-4'-n-alkyloxythiobenzoates (Figure 1), denoted in literature as $\bar{n}$S5 or nOS5. In this paper, we use the abbreviation nOS5. The nOS5 compounds with n = 4-6 exhibit only the nematic phase, the homologue with n = 7 shows the nematic and smectic C phase, while for n = 8-12 there is the nematic phase and polymorphism of the smectic phases [11-32]. There are many publications reporting the X-ray diffraction (XRD) results for the nOS5 compounds, presenting the structural studies of the N, SmA and SmC phases [16,21,25-29], crystal-like smectic phases [12,15,24,28] and crystal phases [29-32]. However, the overall results are still incomplete and do not enable the full structural characterization of the nOS5 compounds.

This study involves three nOS5 homologues with n = 9, 10, 11. They are reported to exhibit N, SmA, SmC, SmJ and SmE phases [12-23]. In the early works, the SmB phase [12] and "tilted SmB phase" (probably meant to be SmJ or SmG) [13] are also mentioned. The published XRD results for these homologues support the information about the N, SmA, SmC, SmB phases for n = 9 [12,16], N, SmA, SmC, SmJ phases for n = 10 [15,16,21] and N, SmA phases for n = 11 [16]. The aim of our work is to present explicitly the XRD patterns of the liquid crystalline phases for n = 9, 10, 11 and to determine the structural parameters for different $C_nH_{2n+1}$ chain lengths as a function of temperature. Wherever possible, we attempt to characterize the metastable, crystal-like smectic phases. Structural studies are supported by polarizing optical microscopy (POM) and differential scanning calorimetry (DSC) results.

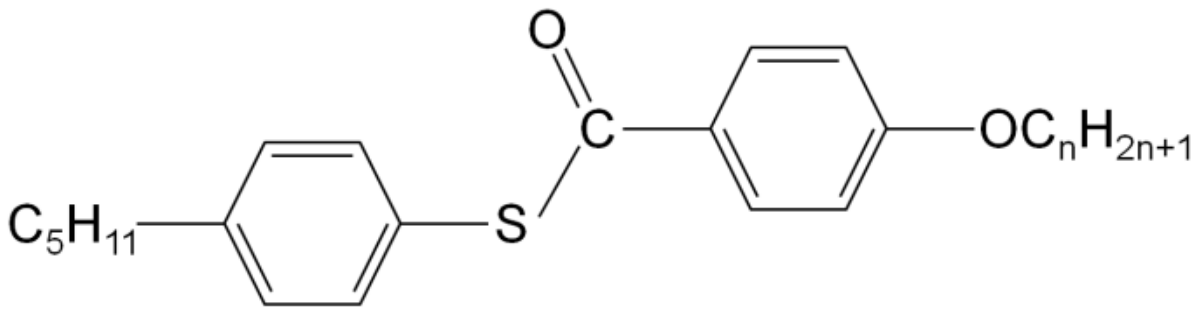

Figure 1. General formula of 4-pentylphenyl-4'-n-alkyloxythiobenzoates. In this paper, the homologues with n = 9, 10, 11 are investigated.

## 2. Experimental details

The nOS5 homologues with n = 9, 10, 11 were synthesized according to the method presented elsewhere [11].

The samples for POM measurements were placed between two glass slides without aligning layers. The texture observations were performed during cooling and heating with the 5 K/min rate in the 263-373 K range using the Leica DM2700 P microscope equipped with the Linkam temperature attachment. The TOApy program [33,34] was used to calculate the average luminance of each texture.

The samples for DSC measurements, weighting 3.16, 3.19 and 3.10 mg for n = 9, 10, 11, respectively, were sealed within aluminium pans. The DSC thermograms were collected in two cycles of heating and cooling with the 5 K/min rate in the 263-373 K range using the PerkinElmer DSC 8000 calorimeter. The calibration was based on the melting points and melting enthalpies of indium and water, the uncertainties of the phase transition temperatures and enthalpy changes are 0.5 K and 0.09 kJ/mol. The data were analysed with the PerkinElmer software and OriginPro.

The XRD patterns were collected in the Bragg-Brentano geometry for the flat samples in the holders of the 13 mm × 10 mm × 0.2 mm size. Two types of samples were used: pure nOS5 compounds and nOS5 mixed with powder $Al_2O_3$. The latter were used to reduce the effect of the preferred orientation, because the smectic layers tend to align parallel to the surface of the flat sample holder. The measurements in the 2θ = 2-30° or 2-8° ranges, for temperatures between 263 K and 373 K, were performed with the CuKα radiation using the X'Pert PRO (PANalytical) diffractometer with the TTK-450 (Anton Paar) temperature attachment. The WinPLOTR [35], FullProf [36] and OriginPro programs were used to analyse the XRD patterns. The molecular models of 9OS5, 10OS5 and 11OS5, used to estimate the molecular length, were optimized in Gaussian 16 [37] with the semi-empirical PM7 method (MOPAC version) [38] and visualized in Avogadro [39].

## 3. Results and discussion

### *3.1. Phase sequence for the 5 K/min cooling/heating rate investigated by POM and DSC*

The selected POM textures of nOS5 with n = 9, 10, 11 are presented in Supporting Information file, in Figures S1-S3. The textures were collected each 1 K and for each of them, the average luminance was calculated to visualize the phase transitions (Figure 2). The DSC curves are shown in Figure 3a-c, and the summary of the phase transitions of nOS5 is presented in Figure 3d and Table 1. Since not all crystal-like smectic phases of nOS5 have been identified to date by the X-ray diffraction, in our discussion we prefer to use the more general notations SmX for the smectic phases with the hexagonal in-planar order (which may be SmB, SmG or SmJ) and SmY for the smectic phases with the herring-bone in-planar order (which may be SmE, SmH or SmK). The crystal phases are denoted as Cr and numbered in an order of decreasing temperature, if necessary.

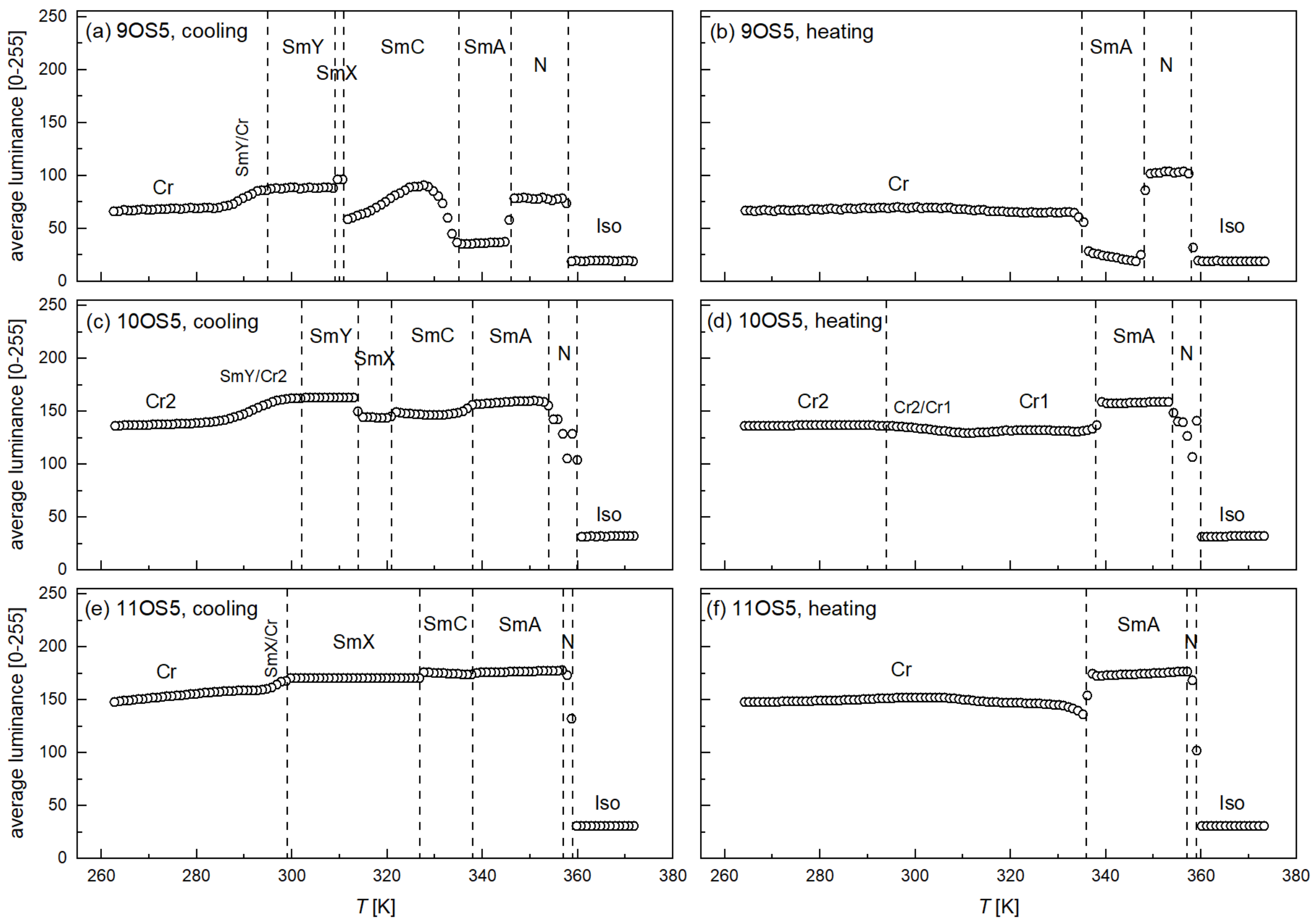

Figure 2. Average luminance as a function of temperature obtained for the POM textures of the nOS5 homologues with n = 9 (a,b), 10 (c,d) and 11 (e,f). The textures were collected during cooling (a,c,e) and heating (b,d,f) with the 5 K/min rate.

During cooling, the POM results indicate the sequence of mesophases N → SmA → SmC → SmX for all investigated compounds. The SmA → SmC transition is not clearly visible in the DSC curves, indicating that it is the 2$^{nd}$ order transition. In lower temperatures, POM observations enable detection of SmY for n = 9, 10, while DSC method reveal additionally the SmY' phase for n = 9 and SmY', SmY phases for n = 11. As the length of the $C_nH_{2n+1}$ chain increases, the temperature ranges of the N and SmC phases decrease and of the SmA and SmX phases increase. The enthalpy changes $\Delta H$ of the Iso ↔ N and N ↔ SmA transitions increase also with increasing n. The homologues with n = 9, 11 show one crystal phase in the investigated temperature range. For 10OS5, the cold crystallization from the metastable Cr2 phase to the Cr1 phase is observed during heating. 10OS5 has also the highest melting temperature and the largest enthalpy of melting. During heating, only SmA and N are observed among the liquid crystalline phases. Although the SmA → SmC transition temperature is slightly higher than the melting temperature of each homologue (Figure 3d), the SmC phase was not detected during heating. If it is present, it may occur only in a very narrow temperature range. Structural characterization of the liquid crystalline phases is discussed in the next section.

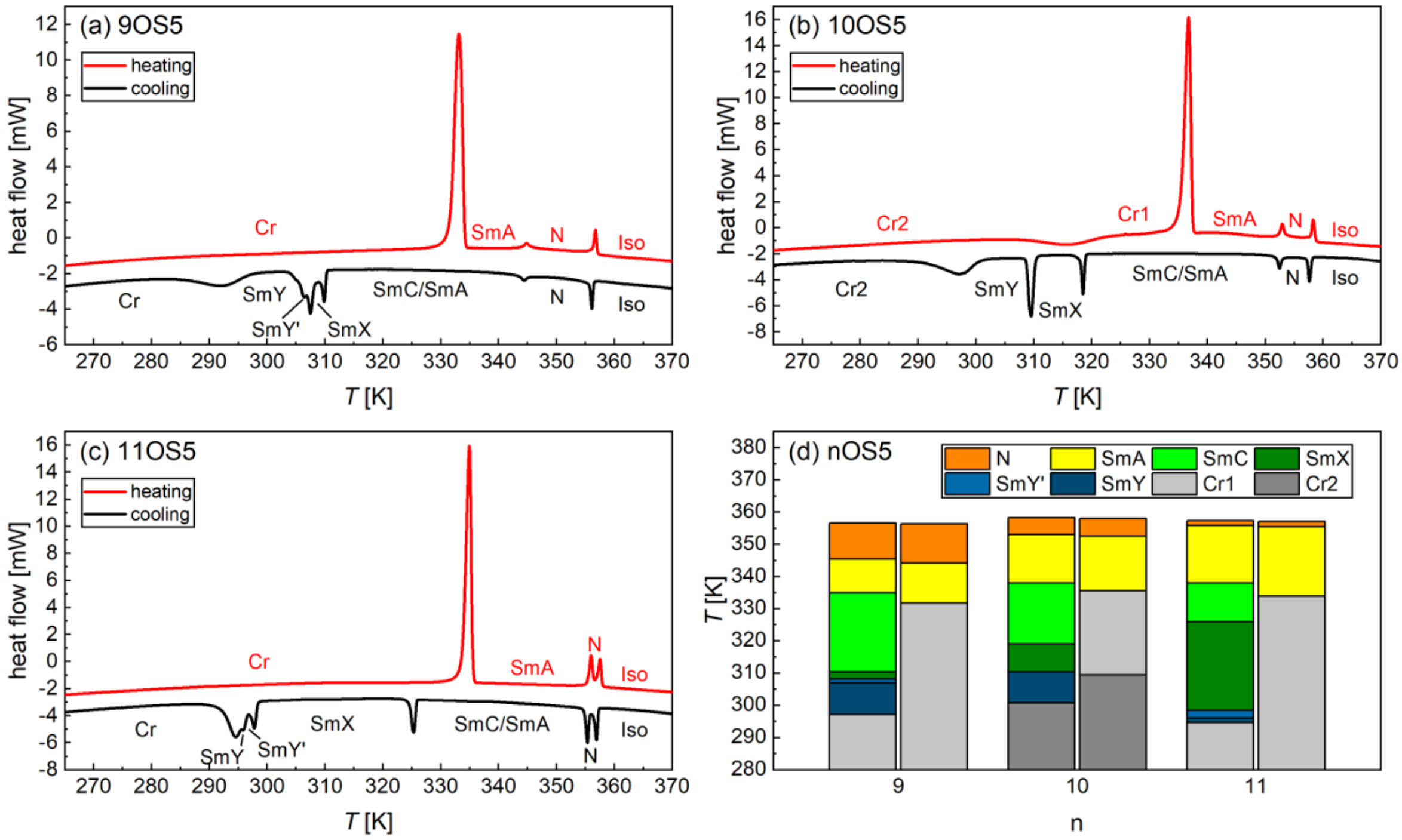

Figure 3. DSC thermograms of the nOS5 homologues with n = 9 (a), 10 (b) and 11 (c) collected during the second heating and second cooling with the 5 K/min rate. Panel (d) shows the summary of the DSC results for all compounds. The SmA → SmC transition temperature was obtained based on POM observations. The left and right columns correspond to the transitions during cooling and heating, respectively.

Table 1. Phase transition temperatures $T_{onset}$ and $T_{peak}$, and enthalpy changes $\Delta H$ obtained for nOS5 (n = 9, 10, 11) by DSC for the 5 K/min rate. The transition temperatures $T_{POM}$ determined by POM are shown for comparison. The minus sign before $\Delta H$ indicates an exothermic transition.

| | $T_{POM}$ [K] | $T_{onset}$ [K] | $T_{peak}$ [K] | $\Delta H$ [kJ/mol] |
|---|---|---|---|---|
| | | 9OS5, cooling | | |
| Iso → N | 358 | 356.6 | 356.2 | −1.6 |
| N → SmA | 346 | 345.4 | 344.6 | −0.8 |
| SmA → SmC | 335 | - | - | - |
| SmC → SmX | 311 | 310.4 | 310.0 | |
| SmX → SmY’ | 309 | 308.2 | 307.6 | |
| SmY’ → SmY | - | 306.9 | 306.5 | −10.6[a] |
| SmY → Cr | 295 | 297.2 | 292.2 | −8.1 |
| | | 9OS5, heating | | |
| Cr → SmA | 335 | 331.8 | 333.3 | 32.8 |
| SmA → N | 348 | 344.2 | 345.0 | 0.7 |
| N → Iso | 358 | 356.4 | 356.9 | 1.4 |
| | | 10OS5, cooling | | |
| Iso → N | 360 | 358.2 | 357.8 | −1.8 |
| N → SmA | 354 | 353.1 | 352.6 | −1.3 |
| SmA → SmC | 338 | - | - | - |
| SmC → SmX | 321 | 319.1 | 318.6 | −3.4 |
| SmX → SmY | 314 | 310.4 | 309.7 | −7.6 |
| SmY → Cr2 | 302 | 300.8 | 297.3 | −12.1 |
| | | 10OS5, heating | | |
| Cr2 → Cr1 | 294 | 309.5 | 316.1 | −9.1 |
| Cr1 → SmA | 338 | 335.6 | 336.9 | 38.1 |
| SmA → N | 354 | 352.5 | 353.1 | 1.6 |
| N → Iso | 360 | 358.0 | 358.4 | 1.5 |
| | | 11OS5, cooling | | |
| Iso → N | 359 | 357.4 | 355.5 | −2.8 |
| N → SmA | 357 | 355.9 | 355.5 | −3.0 |
| SmA → SmC | 338 | - | - | - |
| SmC → SmX | 327 | 326.0 | 325.5 | −3.6 |
| SmX → SmY’ | 299 | 298.5 | 298.0 | |
| SmY’ → SmY | - | - | 296.0 | |
| SmY → Cr | - | - | 294.7 | −19.5[b] |
| | | 11OS5, heating | | |
| Cr → SmA | 336 | 333.9 | 335.1 | 35.5 |
| SmA → N | 357 | 355.5 | 356.1 | 3.1 |
| N → Iso | 359 | 357.1 | 357.7 | 2.7 |

[a] summed enthalpy changes for SmC → SmX → SmY’ → SmY

[b] summed enthalpy changes for SmX → SmY’ → SmY → Cr

### *3.2. Structural studies by XRD*

The XRD patterns of nOS5 were collected using various temperature treatment: (1) upon heating of the pristine samples up to 373 K – 1st heating, (2) on cooling from the isotropic liquid to 298 K – slow cooling, (3) after cooling the sample with the 5 K/min rate directly to a selected temperature – fast cooling, and (4) upon heating after cooling the sample with the 5 K/min rate directly from the isotropic liquid to 263 K – heating after fast cooling. There were two reasons for applying various temperature programs: at first, the crystal phase of the pristine sample may be different than the crystal phase formed from the melt and at second, measurements performed after fast cooling facilitate investigation of the metastable liquid crystalline phases. The selected XRD patterns are presented in Figure 4.

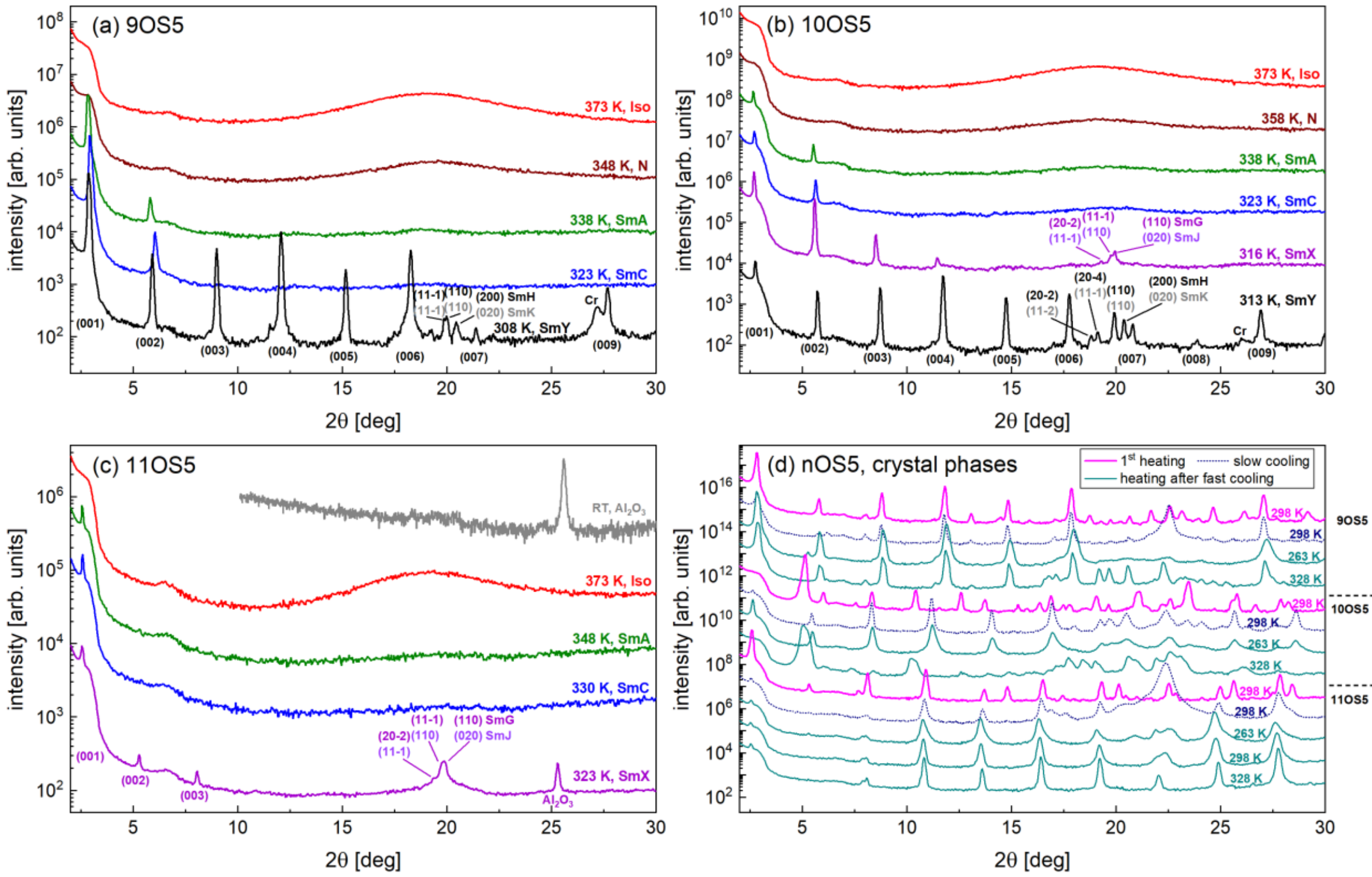

Figure 4. XRD patterns of the isotropic liquid and liquid crystalline phases of 9OS5 (a), 10OS5 (b), 11OS5 (c), and of the crystal phases of all investigated nOS5 homologues (d). The wide maximum at 2θ ≈ 7° is the background contribution. The pattern of $Al_2O_3$, used to reduce the preferred orientation in some samples, is shown in panel (c).

Both in the crystal and smectic phases, the low-angle diffraction peak is observed. In the smectic phases, this peak indicates the layer order and its position corresponds to the smectic layer spacing $d_{001}$. In the crystal phases, the peak's position is similar as in the smectic phases, therefore these phases are likely characterized also by the lamellar order. Usually not only the 1st order peak, but also higher harmonics are visible in the XRD patterns and they were used in calculation of $d_{001}$. The Bragg equation [40], which connects the $d_{001}$ value, the peak's position $\theta_l$ (corrected by the systematic shift $\theta_0$), the X-ray wavelength $\lambda$ and the order of the peak $l$ = 1, 2, 3…, is:

$$l\lambda = 2d_{001}\sin(\theta_l - \theta_0), \tag{1}$$

which can be rearranged to the form:

$$\theta_l = \theta_0 + \operatorname{asin}\left(\frac{l\lambda}{d_{001}}\right). \tag{2}$$

If three or more harmonics are visible in the XRD pattern, the (2) formula can be fitted to the plot of the peaks' positions vs. number of harmonics. This enables simultaneous determination of the layer spacing $d_{001}$ and the systematic shift $\theta_0$. The exemplary fit is presented in Figure S4 in SI. If only two harmonics were visible, the fitting of the (2) formula was performed with a fixed $\theta_0$ parameter, and if only 1$^{st}$ order peak was observed, the layer spacing was calculated from its position corrected by $\theta_0$. In each case, if $\theta_0$ could not be determined for a given temperature, the average $\theta_0$ value obtained for a crystal phase was used and its uncertainty was afterwards included in the total uncertainty of $d_{001}$. The obtained layer spacings are presented in Figure 5. For clarity, the results from cooling are discussed, if not stated otherwise.

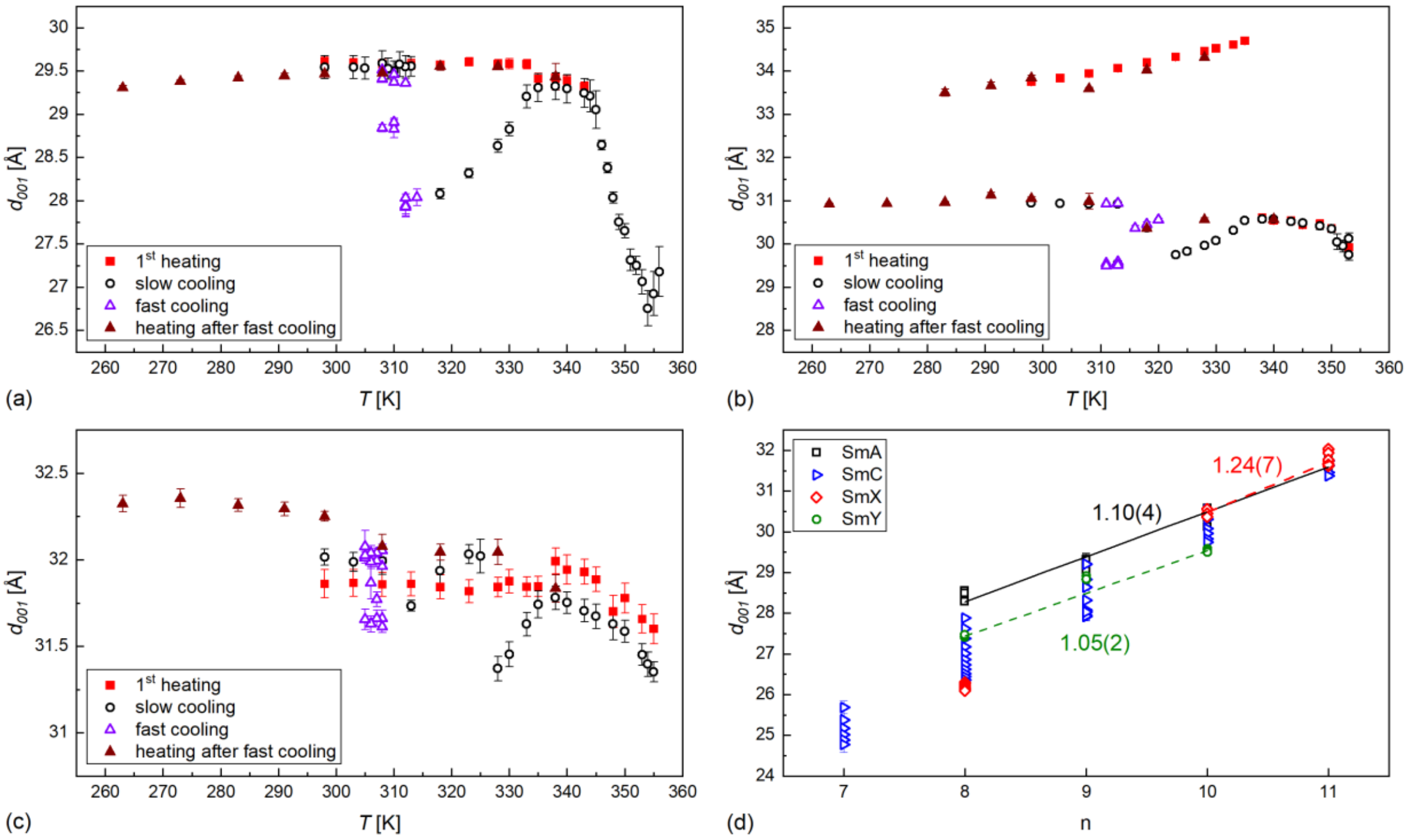

Figure 5. Layer spacing in the smectic phases and crystal phases of the nOS5 homologues with n = 9 (a), 10 (b) and 11 (c) as a function of temperature. Panel (d) shows the layer spacing in different smectic phases, obtained on cooling, as a function of the $C_nH_{2n+1}$ chain length. The layer spacings for n = 7 and 8 were taken from Refs. [29] and [28], respectively.

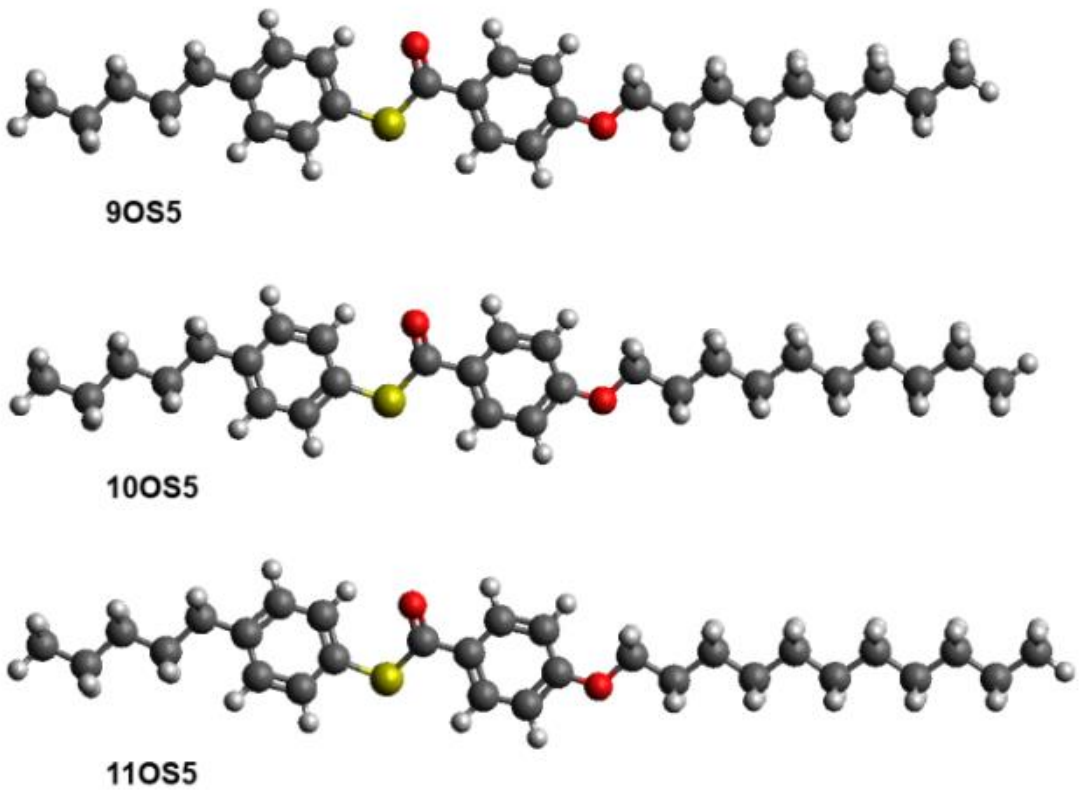

Figure 6. Models of the nOS5 molecules optimized by the PM7 method. The molecular lengths, calculated as the maximal distances between the hydrogen atoms plus non-bonded contact distance for hydrogens (2.2 Å [41]), are equal to 32.7 Å, 33.9 Å and 35.1 Å for n = 9, 10, 11, respectively.

Only for 9OS5, the low-angle maximum is visible in the nematic phase. The distance $d$ in this phase increases with decreasing temperature from 26.8-27.2 Å in 354-356 K to 28.6 Å in 346 K (Figure 5a), which arises from the increasing orientational order parameter [2,12]. In the SmA phase, observed in the 335-345 K range, the layer spacing equals 29.1-29.3 Å. After transition to the SmC phase, $d$ decreases from 29.2 Å in 333 K to 27.9 Å in 312 K. In the 308-310 K range, $d$ = 28.8-28.9 Å and it is interpreted as transition to the SmY phase. The SmX phase was not detected in the XRD data for 9OS5. After crystallization, $d$ = 29.4-29.6 Å, in agreement with 29.3-29.6 Å for the pristine sample on heating.

In the SmA phase of 10OS5, observed in the 338-353 K range, the layer spacing equals 30.1-30.6 Å (Figure 5b). In the SmC phase, $d$ decreases from 30.5 Å in 335 K to 29.8 Å in 323 K. In the SmX phase, $d$ increases to 30.4-30.6 Å in 316-320 K, and in the SmY phase, $d$ = 29.5-29.6 Å in 311-313 K. After crystallization, $d$ increases to 30.9-31.0 Å. In the pristine sample on heating, the spacing is larger and equal to 33.8-34.7 Å. Upon heating after fast cooling, the Cr1 → Cr2 transition is confirmed: in the low-temperature phase, $d$ = 30.9-31.1 Å, and in the high-temperature phase, $d$ = 33.2-33.8 Å, similarly as in the pristine sample.

11OS5 shows the smallest changes in the layer spacing (Figure 5c). In the SmA phase in the 338-350 K range, $d$ = 31.6-31.8 Å. The SmC phase is observed in the 328-335 K range, with $d$ = 31.4-31.7 Å. After transition to the SmX phase, the layer spacing increases to 32.0 Å in 325 K and decreases further with decreasing temperature to 31.6-31.7 Å in 305-306 K. The SmY' and SmY phases were not detected. After crystallization, $d$ increases slightly to 32.0-32.1 Å. In the pristine sample on heating, $d$ is smaller and equal to 31.8-31.9 Å, while on heating after fast cooling, $d$ is equal to 32.0-32.4 Å.

Figure 5d shows the layer spacing in various smectic phases in nOS5, including also the earlier results for m = 7, 8 from [28,29]. The layer spacing in the SmA and SmX phases has a slope of 1.10(4) Å for n = 8-11 and 1.24(7) Å for n = 10, 11 per a $CH_2$ group, respectively. It corresponds approximately to the increase of the molecular length by 1.2 Å after addition of one $CH_2$ group (Figure 6). These results indicate that the smectic phases of nOS5 are of the simplest, monolayer type and the molecular dimers are not formed (as it would lead to a larger slope [42]). The exception is the SmX phase of 8OS5, where the layer spacing is lower than extrapolated from the results for n = 10, 11. Thus, the SmX phase of 8OS5 has a different structure than for longer homologues, but it is certainly a tilted hexagonal phase (in Ref. [28], it was referred to as SmJ based on an earlier source [43]). For the SmY phase, the slope between n = 8 and 10 is 1.05(2) Å per a $CH_2$ group, while for n = 9 the layer spacing is larger, therefore SmY can be a different phase for 9OS5 than for n = 8, 10. In Ref. [28], the SmY phase of 8OS5 was identified as orthogonal SmE due to the increase of the layer spacing at the SmX → SmY transition. However, as it is further discussed, the SmY phase of 10OS5 is a tilted herring-bone smectic, and the same is probably for 8OS5.

In the XRD patterns of the crystal-like smectic phases (Figure 4a-c), the sharp diffraction peaks are observed at higher 2θ angles and they enable determination of the unit cell parameters (Table 2). The SmB phase has a hexagonal unit cell with $a = b \neq c$, $\alpha = \beta = 90°$ and $\gamma = 120°$. The SmE phase has an orthorhombic unit cell with the parameters $a \neq b \neq c$, $\alpha = \beta = \gamma = 90°$, in a convention $b < a < c$. In the tilted hexatic, hexagonal and herring-bone phases, the positional order of molecules can be described by the monoclinic unit cell (in the hexatic phases it applies only to the local order) with the parameters $a \neq b \neq c$, $\alpha = \gamma = 90°$ and $\beta \neq 90°$. For the SmF, SmG, SmH phases there is a relationship $b < a < c$, while for the SmI, SmJ, SmK phases, it is $a < b < c$ [1,44,45]. The tilt angle of molecules is obtained as $\Theta = \beta - 90°$.

There are three clearly visible peaks attributed to the SmY phase of 9OS5, which are indexed as $11\bar{1}$, $110$, $200$ if one assumes the SmH phase, and $11\bar{1}$, $110$, $020$ if one assumes the SmK phase. The obtained unit cell of SmH is characterized by the $a/b$ ratio very close to $\sqrt{3} \approx 1.732$ for an ideal hexagonal order, and the tilt angle $\Theta \approx 13°$. For the unit cell of SmK, $b/a < \sqrt{3}$ and $\Theta \approx 8°$. Although two first peaks have very close positions, diffraction patterns of SmY cannot be indexed as the orthogonal SmE phase [46].

For the SmX phase registered for 10OS5 and 11OS5, up to three diffraction peaks describing the ordering within the smectic layers are visible. With an assumption that SmX is the SmG phase, these peaks can be indexed as $20\bar{2}$, $11\bar{1}$ and $110$, where the strongest peak in the 2θ range around 20° is indexed as $110$ [44]. Together with the $001$ peak and its higher harmonics related to the smectic layer spacing, they are used to calculate the monoclinic unit cell parameters, using a method described in details in SI. If one assumes that SmX is the SmJ phase, the peaks are indexed as $11\bar{1}$, $110$ and $002$, where $002$ is attributed to the strongest peak [45]. The $a/b$ ratio is above $\sqrt{3}$ for SmG and below $\sqrt{3}$ for SmJ. The tilt angle $\Theta \approx$ 15-16° in both cases. The SmX phase of 11OS5 is observed in the XRD patterns down to 305 K, but only for 323 K the diffraction peaks are visible enough to determine the unit cell size.

Four diffraction peaks were registered for the SmY phase 10OS5. If one assumes the transition to the SmH phase, the diffraction peaks can be indexed as $20\bar{2}$, $20\bar{4}$, $110$ and $200$. The unit cell parameters are obtained from the positions of the $00l$, $20\bar{2}$, $110$, $200$ peaks, and the position of the $20\bar{4}$ peak is then correctly reproduced. If SmY is the SmK phase, then the diffraction peaks are indexed as $11\bar{2}$, $11\bar{1}$, $110$, $020$. The unit cell parameters are calculated based on the $00l$, $11\bar{1}$, $110$, $020$ peaks, and the position of the $11\bar{2}$ peak is correctly reproduced. The tilt angle $\Theta$ = 22.6-23.6° is higher than in SmX, which corresponds to the decrease in the layer spacing at the SmX → SmY transition. The relationship of the in-planar distances is $a/b > \sqrt{3}$ for SmH and $b/a < \sqrt{3}$ for SmK.

In all unit cells, the $c$ parameter is slightly shorter than the calculated molecular length. The molecular models presented in Figure 6 are in the all-trans conformation, which was suggested by the experimental crystal structures of shorter nOS5 homologues [29-32], and the maximal molecular lengths are calculated. The XRD results for SmX and SmY show that in these phases, the terminal chains may be not in the all-trans conformation, but conformationally disordered, as it is assumed for the SmE phase [47]. In the crystal-like smectic phases, the unit cell contains two molecules [1]. Using the unit cell volume determined by XRD, one can obtain the mass density. For the SmJ phase, $\rho$ = 1.00-1.02 g/cm$^3$, which is slightly smaller than for the SmH and SmK phases, 1.03-1.04 g/cm$^3$. For comparison, the densities of the shorter nOS5 compounds (n = 4-7) in their crystal phases are in the 1.104-1.177 g/cm$^3$ range [29-32].

Based on the presented powder diffraction results, we cannot distinguish SmG from SmJ, and SmH from SmK, because an assumption of each phase from the pair leads to reasonable results. However, the XRD method allows us to confirm that SmX for n = 10, 11 is the tilted hexagonal phase, and SmY for n = 9, 10 is the tilted herring-bone phase. The SmX, SmY' phases of 9OS5 and SmY', SmY phases of 11OS5 were not detected in the diffraction patterns.

Table 2. Unit cell parameters of the crystal-like smectic phases of nOS5.

| phase | assumption | $T$ [K] | $a$ [Å] | $b$ [Å] | $c$ [Å] | $\beta$ [deg] | $a/b$ or $b/a$ | $V$ [Å$^3$] | $\rho$ [g/cm$^3$] |
|---|---|---|---|---|---|---|---|---|---|
| 9OS5 | | | | | | | | | |
| SmY | SmH | 310 | 8.85(3) | 5.100(4) | 29.6(2) | 103.2(6) | 1.735(5) | 1301(8) | 1.090(7) |
| | | 308 | 8.84(3) | 5.112(5) | 29.7(1) | 103.5(7) | 1.730(5) | 1304(7) | 1.087(6) |
| | SmK | 310 | 5.15(2) | 8.61(2) | 29.1(2) | 98(1) | 1.673(7) | 1278(8) | 1.109(7) |
| | | 308 | 5.162(9) | 8.600(8) | 29.12(6) | 98.0(6) | 1.666(4) | 1280(4) | 1.107(4) |
| 10OS5 | | | | | | | | | |
| SmX | SmG | 320 | 9.12(2) | 5.112(3) | 31.71(7) | 105.5(4) | 1.784(4) | 1425(5) | 1.027(4) |
| | | 318 | 9.11(4) | 5.092(5) | 31.5(2) | 105.7(7) | 1.788(7) | 1408(9) | 1.039(7) |
| | | 316 | 9.11(4) | 5.093(5) | 31.5(2) | 105.6(8) | 1.788(8) | 1408(10) | 1.039(7) |
| | SmJ | 320 | 5.344(9) | 8.838(3) | 31.62(6) | 104.9(3) | 1.654(3) | 1443(4) | 1.014(3) |
| | | 318 | 5.35(2) | 8.806(6) | 31.5(1) | 105.7(6) | 1.647(5) | 1430(8) | 1.023(6) |
| | | 316 | 5.35(2) | 8.808(6) | 31.5(1) | 105.6(5) | 1.648(5) | 1430(7) | 1.024(5) |
| SmY | SmH | 313 | 9.37(1) | 5.125(3) | 32.16(4) | 113.2(2) | 1.829(3) | 1419(3) | 1.031(2) |
| | | 311 | 9.38(1) | 5.126(4) | 32.24(4) | 113.6(2) | 1.830(3) | 1421(3) | 1.030(2) |
| | SmK | 313 | 5.55(2) | 8.612(4) | 32.03(8) | 112.7(3) | 1.550(4) | 1414(6) | 1.035(4) |
| | | 311 | 5.55(2) | 8.587(4) | 32.00(8) | 112.6(3) | 1.548(4) | 1410(6) | 1.038(5) |
| 11OS5 | | | | | | | | | |
| SmX | SmG | 323 | 9.08(8) | 5.11(2) | 33.2(4) | 104(2) | 1.78(2) | 1493(25) | 1.01(2) |
| | SmJ | 323 | 5.30(3) | 8.84(2) | 33.1(3) | 103(2) | 1.67(1) | 1508(16) | 1.00(1) |

In the nematic and SmA phases, the molecules are not aligned perfectly but they are tilted in random directions and only the average tilt angle is zero. The orientational order parameter is defined as [2,12]:

$$S = \langle (3\cos^2\vartheta - 1)/2 \rangle, \quad (3)$$

where $\vartheta$ is an angle between the long molecular axis and director (the resultant direction of the long axes of molecules), and angle brackets indicate averaging over molecules. In the isotropic liquid phase, without any orientational order, $S$ equals zero, while in the nematic and SmA phase it is usually 0.3-0.8 [2,12,22,48,49]. By comparison of the layer spacing $d_{001}$ and the molecular length $L$, one can obtain the average absolute tilt as $\cos\vartheta = d_{001}/L$ and insert it into the (3) formula to estimate the orientational order parameter. The $L$ values determined by the PM7 calculations were applied. The $S$ values in the SmA are equal to 0.70-0.75 for all homologues (Figure 7a). The low-angle diffuse maximum in the XRD patterns of the nematic phase was observed only for 9OS5, which enabled determination of the order parameter also at higher temperatures. The $S$ value in the N phase of 9OS5 equals 0.5-0.54 in a few degree below the Iso → N transition and it increases on cooling to 0.65 before the N → SmA transition. The $S$ parameter obtained for 10OS5 based on the measurements of absorption and fluorescence reaches 0.75 in the SmA phase [22] and it agrees with the XRD results.

In the tilted smectic phases, the director makes an angle $\Theta$ with the layer normal. For the rod-like molecules, the tilt angle can be determined from the formula $\cos\Theta = d_{001}/L$. Since there may be still some orientational disorder in the SmC phase (i.e. the molecules can be locally tilted from the director), the $\Theta$ angle was calculated by inserting as $L$ the largest layer spacing in the SmA phase for each homologue. For the crystal-like SmX and SmY phases, the $\Theta$ values from Table 2 were used ($\Theta = \beta - 90°$). The tilt angle increases smoothly on cooling in the SmC and SmX phases of 10OS5 and 11OS5, without any anomalies at the SmC → SmX transition (Figure 7b). The tilt angle shows similar dependence on temperature for all homologues and reaches 14-18°, which is smaller than for the SmC phase of 8OS5, where $\Theta$ obtained from the XRD patterns reaches 19-22° [27,28]. Only the transition to the SmY phase is connected with an abrupt increase of $\Theta$ to 23-24° for 10OS5 and decrease to 8° or 13° for 9OS5.

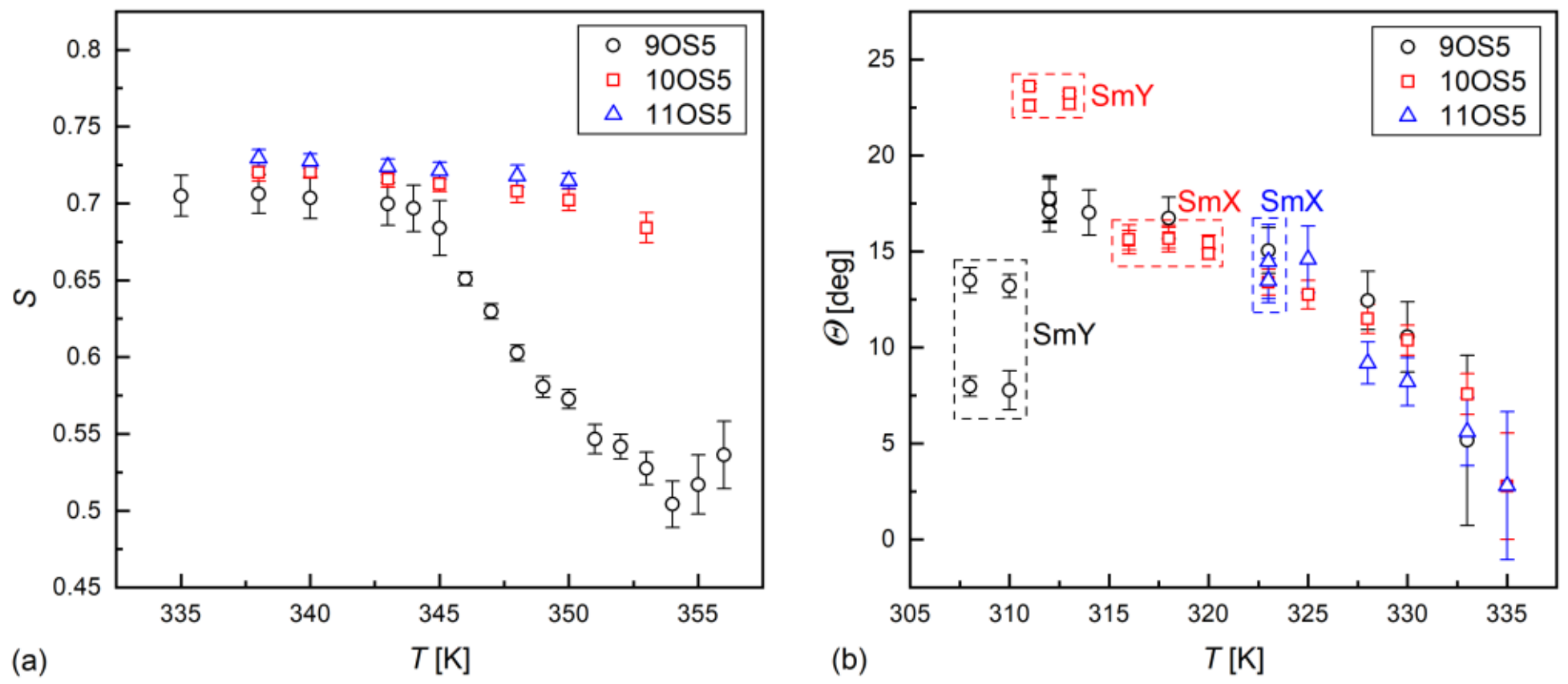

Figure 7. Orientational order parameter in the N (for n = 9 only) and SmA phases (a), and the tilt angle in the SmC, SmX and SmY phases of nOS5 (b) determined from the XRD patterns.

The diffuse maximum at 2θ ≈ 18° in the diffraction patterns of the isotropic liquid, nematic, SmA and SmC phase corresponds to the positional short-range order, which parameters are the correlation length $\xi$ and the average distance between molecules $w$. The shape of this maximum in the space of the scattering vector $q = 4\pi \sin\theta / \lambda$ is the Lorentz peak function [1,50]:

$$I(q) = \frac{A}{1+\xi^2(q-q_0)^2} + Bq + C, \quad (4)$$

where the maximum's position $q_0 = 2\pi/w$, $A$ is the height of maximum and the $B$, $C$ parameters are added to fit the linear background. The representative XRD patterns at higher $q$ region are shown for 9OS5 in Figure 8a. The nOS5 samples mixed with powder $Al_2O_3$ usually enabled better visibility of the diffuse peak by reducing the effect of the preferred orientation, therefore these results were used for fitting of the (4) formula. For 10OS5, the patterns obtained upon the 1st heating were also included, as the diffuse maximum was well-visible in them. The average intermolecular distance $w$ equals 4.5-4.7 Å, without significant changes at the phase transitions, and is comparable with the width of molecules. The correlation length in the isotropic liquid phase equals 3.9-4.6 Å, indicating only the next-neighbour interactions. In the nematic and both smectic phases, the increasing $\xi$ values with decreasing temperature are obtained and they reach maximally 9.6, 7.9 and 9.2 Å for n = 9, 10, 11, respectively (Figure 8b). The correlation length is comparable with the doubled $w$, indicating the next-neighbors interactions within the smectic layers. The correlation length determined previously for 8OS5 reaches maximally 9.0 Å in the SmC phase [28]; the maximal values reported for other mesogens in the SmA or SmC phases are 7.1-7.6 Å [51], 15.8 Å [52], 22 Å [53], 23-24.5 Å [54], 25.5-27 Å [55]. In references [54,55], the correlation length is calculated as $\xi = 2\pi/FWHM$, where $FWHM$ is the full-width at half-maximum of the diffuse peak. However, if one uses the (4) formula, given in [1,50], it implicates $\xi = 2/FWHM$ and for purpose of comparison, the $\xi$ values from [54,55] should be divided by $\pi$, giving 7.3-8.6 Å.

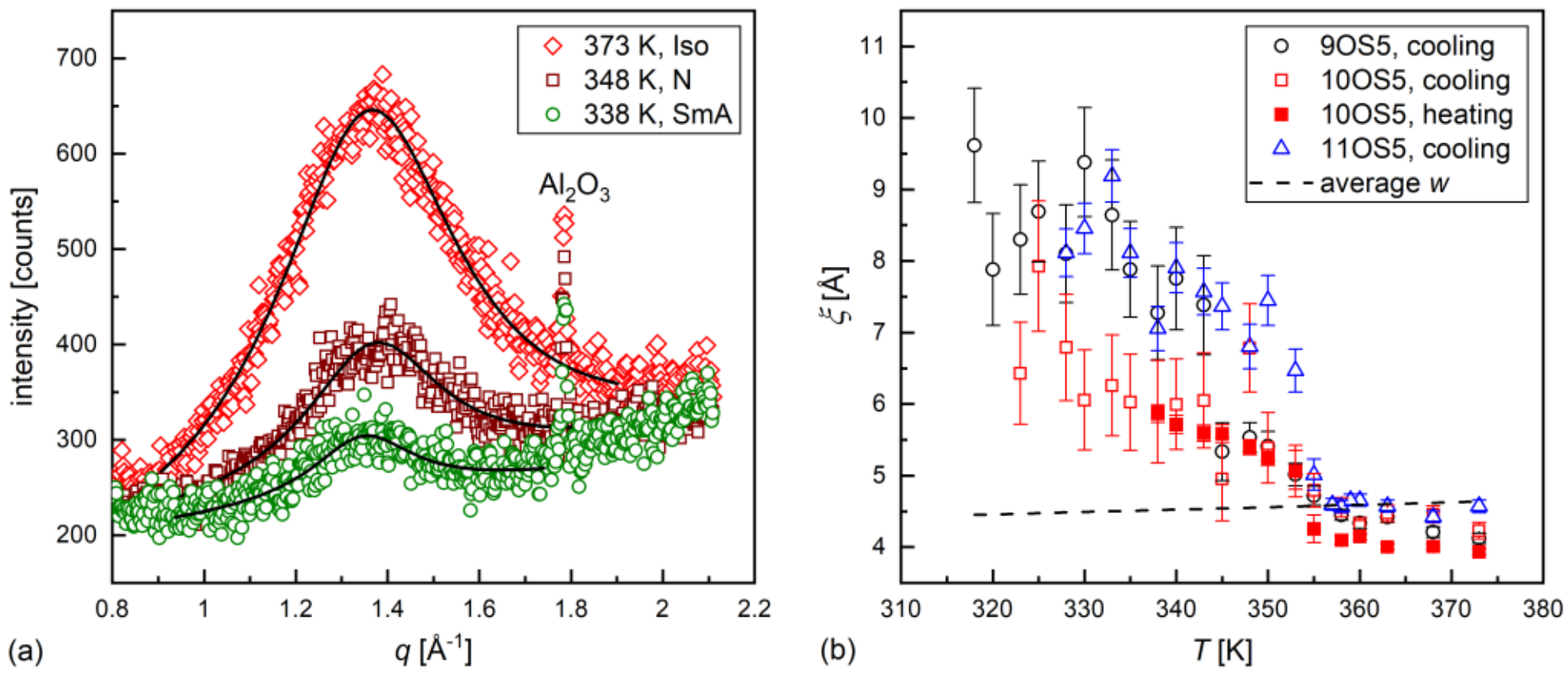

Figure 8. Diffuse maximum in the diffraction patterns of 9OS5 mixed with $Al_2O_3$ with the fitted (4) formula (a) and the correlation length and average intermolecular distance obtained for all investigated homologues (b). The sharp peak from $Al_2O_3$ visible in (a) was excluded from fitting.

## 4. Summary and conclusions

Three compounds from the homologous series of 4-pentylphenyl-4'-n-alkyloxythiobenzoates (nOS5), differing by the length of the $C_nH_{2n+1}$ terminal chain (n = 9, 10, 11), were investigated by polarizing optical microscopy, differential scanning calorimetry, X-ray diffraction and PM7 semi-empirical calculations. All compounds form the enantiotropic nematic and smectic A phases, and monotropic smectic C phase. On further cooling, the monotropic crystal-like tilted smectic phases are observed. The main conclusions in this subject are as follows:

- 9OS5 exhibits the smectic phases denoted as SmX, SmY' and SmY. Based on the XRD patterns, SmY is interpreted as the SmH or SmK phase, with the herring-bone ordering within the smectic layers and a small tilt angle of molecules.
- 10OS5 shows the SmX and SmY phases, where SmX is either SmG or SmJ with the hexagonal order within layers, and SmY is either SmH or SmK.
- 11OS5 exhibits the SmX, SmY' and SmY phases, where SmX can be SmG or SmJ.
- Some of the monotropic phases could not have been studied by the conventional XRD method because they underwent the quick transition to another phase.
- The SmG/SmJ and SmH/SmK pairs differ only by the direction of the molecular tilt and each phase from the pair is in accordance with the corresponding powder diffraction patterns, therefore they cannot be distinguished.

The XRD method enables determination of the tilt angle, which is below 20° for all homologues, except the SmY phase of 10OS5, where it is equal to 23-24°. By inserting the molecular length calculated by the PM7 method, the orientational order parameter is obtained, which is equal to 0.5-0.65 in the nematic phase of 9OS5 and 0.7-0.73 in the smectic A phase of all compounds. The correlation length of the short-range order indicates that the correlation within the smectic layers reaches only the next-nearest neighbours in the smectic A and C phases. The results of the conventional powder XRD method, presented in this study, may be extended in the future by the diffraction experiments of the synchrotron radiation, which can possibly enable more precise identification of the monotropic crystal-like mesophases of nOS5.

**Acknowledgement:** We thank Assoc. Prof. Ewa Juszyńska-Gałązka from Institute of Nuclear Physics Polish Academy of Sciences for the discussion regarding the data presentation. The PerkinElmer DSC 8000 calorimeter was purchased thanks to the financial support of the European Regional Development Fund in the framework of the Polish Innovation Economy Operational Program (contract no. POIG.02.01.00-12-023/08). This research was supported in part by the Excellence Initiative – Research University Program at the Jagiellonian University in Kraków. We gratefully acknowledge Poland's high-performance Infrastructure PLGrid Academic Computer Centre Cyfronet AGH for providing computer facilities and support within computational grant.

**Conflicts of interest statement:** There are no conflicts to declare.

**Authors' contributions:**

A. Deptuch – conceptualization, investigation (PM7, XRD, POM), formal analysis (XRD, POM), writing – original draft.

B. Sęk – investigation (POM), formal analysis (XRD, POM), writing – review and editing.

S. Lalik – investigation (DSC), formal analysis (DSC), writing – review and editing.

M.D. Ossowska-Chruściel – resources (samples' synthesis), writing – review and editing.

J. Chruściel – resources (samples' synthesis), writing – review and editing.

M. Marzec – funding acquisition, resources, writing – review and editing.

**Bibliography:**

[1] D. Demus, J. Goodby, G.W. Gray, H.-W. Spiess, V. Vill (Eds.), Handbook of Liquid Crystals, WILEY-VCH Verlag GmbH, Weinheim 1998.

[2] G. Vertogen, W.H. de Jeu, Thermotropic Liquid Crystals, Fundamentals, Springer-Verlag, Berlin Heidelberg 1988, https://doi.org/10.1007/978-3-642-83133-1.

[3] J. Als-Nielsen, J.D. Litster, R.J. Birgeneau, M. Kaplan, C.R. Safinya, A. Lindegaard-Andersen, S. Mathiesen, Observation of algebraic decay of positional order in a smectic liquid crystal, Phys. Rev. B 22 (1980) 312-320, https://doi.org/10.1103/PhysRevB.22.312.

[4] M. Piwowarczyk, N. Osiecka-Drewniak, M. Gałązka, Z. Galewski, Synthesis, mesogenic and photoisomerization studies of (E)-4-[(4-pentyloxyphenyl)diazenyl]phenyl alkanoates, Phase Trans. 92 (2019) 1066-1076, https://doi.org/10.1080/01411594.2019.1650934.

[5] M. Urbańska, P. Morawiak, M. Senderek, Investigation of the tilt angle and spontaneous polarization of antiferroelectric liquid crystals with a chiral centre based on (S)-(+)-3-octanol, J. Mol. Liq. 328 (2021) 115378, https://doi.org/10.1016/j.molliq.2021.115378.

[6] M.A. Zakaria, M. Alazmi, K.D. Katariya, Y. El Kilany, El S.H. El Ashry, M. Jaremko, M. Hagar, S.Z. Mohammady, Mesomorphic Behaviour and DFT Insight of Arylidene Schiff Base Liquid Crystals and Their Pyridine Impact Investigation, Crystals 11 (2021) 978, https://doi.org/10.3390/cryst11080978.

[7] A. Drzewicz, E. Juszyńska-Gałązka, A. Deptuch, P. Kula, Effect of Alkyl Chain Length on the Phase Situation of Glass-Forming Liquid Crystals, Crystals 12 (2022) 1401, https://doi.org/10.3390/cryst12101401.
[8] K. Strójwąs, R. Dąbrowski, W. Drzewiński, M. Szarek, A. Bubnov, M. Czerwiński, The comparison of self-assembling behaviour of phenyl biphenylcarboxylate and biphenyl benzoate compounds with the different length and shape of chiral terminal chain, J. Mol. Liq. 369 (2023) 120882, https://doi.org/10.1016/j.molliq.2022.120882.
[9] S. Yamada, T. Konno, Development of Donor-π-Acceptor-Type Fluorinated Tolanes as Compact Condensed Phase Luminophores and Applications in Photoluminescent Liquid-Crystalline Molecules, Chem. Rec. 23 (2023) e202300094, https://doi.org/10.1002/tcr.202300094.
[10] M.E. Neubert, B. Ziemnicka-Merchant, M.R. Jirousek, S.J. Laskos, Jr., D. Leonhardt, R. B. Sharma, The Effect of Two Terminal Alkoxy Groups on the Mesomorphic Properties of 4, 4′-Disubstituted Phenylthiobenzoates, Mol. Cryst. Liq. Cryst. 154 (1988) 209-239, https://doi.org/10.1080/00268948808078733.
[11] J. Chruściel, S. Wróbel, H. Kresse, S. Urban, W. Otowski, Dielectric Studies of 4-n-Pentylphenyl-4-Octyloxythiobenzoate, Mol. Cryst. Liq. Cryst. 127 (1985) 57-65, https://doi.org/10.1080/00268948508080831.
[12] A. de Vries, The description of the smectic A and C phases and the smectic A-C phase transition of TCOOB with a diffuse-cone model, J. Chem. Phys. 71 (1979) 25-31, https://doi.org/10.1063/1.438123.
[13] D. Brisbin, R. DeHoff, T.E. Lockhart, D.L. Johnson, Specific Heat near the Nematic-Smectic-A Tricritical Point, Phys. Rev. Lett. 43 (1979) 1171-1174, https://doi.org/10.1103/PhysRevLett.43.1171.
[14] M.E. Neubert, R.E. Cline, M.J. Zawaski, P.J. Wildman, A. Ekachai, The Effect on Mesomorphic Properties of Substituting a Sulfur for the Ether Oxygen Atom in the Ester Linkage of 4-Alkylphenyl-4′-Alkyl or Alkoxybenzoates, Mol. Cryst. Liq. Cryst. 76 (1981) 43-77, https://doi.org/10.1080/00268948108074675.
[15] A.J. Leadbetter, P.A. Tucker, G.W. Gray, A.R. Tajbakhsh, The Phase Behaviour of 4-n-Hexylphenyl 4-n-Tetra-Decyloxybenzthiolate (14S6) and 4-n-Pentylphenyl 4-n-Decyloxybenzthiolate (10S5), Mol. Cryst. Liq. Cryst. Lett. 1 (1985) 19-24, https://doi.org/10.1080/01406566.1985.10766960.
[16] B.M. Ocko, R.J. Birgeneau, J.D. Litster, Crossover to tricritical behavior at the nematic to smectic A transition: An x-ray scattering study, Z. Phys. B Condens. Matter 62 (1986) 487-497, https://doi.org/10.1007/BF01303581.
[17] J. Chruściel, S. Wróbel, H. Kresse, S. Urban, W. Otowski, Odd-even effect in the homologous series of thioesters, Mol. Cryst. Liq. Cryst. 192 (1990) 107-112, https://doi.org/10.1080/00268949008035615.
[18] J. Chruściel, H. Kresse, S. Urban, Megahertz dielectric relaxation process in the nematic and smectic phases of two thiol esters (9S5 and 10S5), Liq. Cryst. 11 (1992) 711-718, https://doi.org/10.1080/02678299208029022.
[19] J. Chruściel, B. Gestblom, M. Makrenek, W. Haase, M. Pfeiffer, S. Wróbel, Molecular dynamics in the nematic, $S_A$ and $S_C$ phases of 8SS and 9S5 as studied by dielectric methods, Liq. Cryst. 14 (1993) 565-572, https://doi.org/10.1080/02678299308027672.
[20] A. Żywociński, S.A. Wieczorek, Critical Exponents for Thermal Expansion and Isothermal Compressibility near the Nematic to Smectic-A, Phase Trans. 101 (1997) 6970-6976, https://doi.org/10.1021/jp971234g.
[21] E. Anesta, G.S. Iannacchione, C.W. Garland, Critical linear thermal expansion in the smectic-A phase near the nematic-smectic phase transition, Phys. Rev. E 70 (2004) 041703, https://doi.org/10.1103/PhysRevE.70.041703.
[22] D. Bauman, A. Zięba, E. Mykowska, Oriental behaviour of some homologues of 4-n-pentyl-phenylthio-4′-n-alkoxybenzoate doped with dichroic dye, Opto-Electron. Rev. 16 (2008) 244-250, https://doi.org/10.2478/s11772-008-0020-5.
[23] J. Chruściel, S. Wojciechowska, M.D. Ossowska-Chruściel, A. Rudzki, S. Zalewski, Phase behaviour of novel liquid crystalline mixtures based on thiobenzoates, Phase Trans. 80 (2007) 615-629, https://doi.org/10.1080/01411590701339807.
[24] P.A.C. Gane, A.J. Leadbetter, Modulated crystal B phases and B-to-G phase transitions in two types of liquid crystalline compound, J. Phys. C: Solid State Phys. 16 (1983) 2059-2067, https://doi.org/10.1088/0022-3719/16/11/009.
[25] C.R. Safinya, M. Kaplan, J. Als-Nielsen, R.J. Birgeneau, D. Davidov, J.D. Litster, D.L. Johnson, M.E. Neubert, High-resolution x-ray study of a smectic-A–smectic-C phase transition, Phys. Rev. B 21 (1980) 4149-4153, https://doi.org/10.1103/PhysRevB.21.4149.

[26] P.S. Clegg, R.J. Birgeneau, S. Park, C.W. Garland, G.S. Iannacchione, R.L. Leheny, M.E. Neubert, High-resolution x-ray study of the nematic–smectic-A and smectic-A–smectic-C transitions in liquid-crystal–aerosil gels, Phys. Rev. E 68 (2003) 031706, https://doi.org/10.1103/PhysRevE.68.031706.

[27] B. Freelon, M. Ramazanoglu, P.J. Chung, R.N. Page, Yuan-Tse Lo, P. Valdivia, C.W. Garland, R.J. Birgeneau, Smectic-A and smectic-C phases and phase transitions in 8S5 liquid-crystal–aerosil gels, Phys. Rev. E 84 (2011) 031705, https://doi.org/10.1103/PhysRevE.84.031705.

[28] A. Deptuch, T. Jaworska-Gołąb, M. Marzec, J. Fitas, K. Nagao, J. Chruściel, M.D. Ossowska-Chruściel, Physical properties of 8OS5 thioester studied by complementary methods, Phase Trans. 90 (2017) 765-772, https://doi.org/10.1080/01411594.2016.1277221.

[29] A. Deptuch, T. Jaworska-Gołąb, J. Kusz, M. Książek, K. Nagao, T. Matsumoto, A. Yamano, M.D. Ossowska-Chruściel, J. Chruściel, M. Marzec, Single crystal X-ray structure determination and temperature-dependent structural studies of the smectogenic compound 7OS5, Acta Cryst. B 76 (2020) 1128-1135, https://doi.org/10.1107/S2052520620014481.

[30] Z. Karczmarzyk, M.D. Ossowska-Chruściel, J. Chruściel, The Crystal and Molecular Structure of 4-n-Pentylphenyl-4′-n-Hexyloxythiobenzoate (6OS5), Mol. Cryst. Liq. Cryst. 357 (2001) 117-125, https://doi.org/10.1080/10587250108028248.

[31] J. Chruściel, B. Pniewska, M.D. Ossowska-Chruściel, The Crystal and Molecular Structure of 4-Pentylphenyl-4′-Pentioxythiobenzoate (5S5), Mol. Cryst. Liq. Cryst. 258 (1995) 325-331, https://doi.org/10.1080/10587259508034572.

[32] M.D. Ossowska-Chruściel, Z. Karczmarzyk, J. Chruściel, The Polymorphism Of 4-N-Pentylphenyl-4″-N-Butyloxythio-Benzoate, (4OS5) In The Crystalline State, Mol. Cryst. Liq. Cryst. 382 (2002) 37-52, https://doi.org/10.1080/713738755.

[33] N. Osiecka, Z. Galewski, M. Massalska-Arodź, TOApy program for the thermooptical analysis of phase transitions, Thermochim. Acta 655 (2017) 106-111, https://doi.org/10.1016/j.tca.2017.06.012.

[34] N. Osiecka-Drewniak, Z. Galewski, E. Juszyńska-Gałązka, Distinguishing the Focal-Conic Fan Texture of Smectic A from the Focal-Conic Fan Texture of Smectic B, Crystals 13 (2023) 1187, https://doi.org/10.3390/cryst13081187.

[35] T. Roisnel, J. Rodriguez-Carvajal, WinPLOTR: A Windows Tool for Powder Diffraction Pattern Analysis, Mater. Sci. Forum, 378-381 (2001) 118-123, https://doi.org/10.4028/www.scientific.net/MSF.378-381.118.

[36] J. Rodríguez-Carvajal, Recent advances in magnetic structure determination by neutron powder diffraction, Phys. B: Cond. Matt. 192 (1993) 55-69, https://doi.org/10.1016/0921-4526(93)90108-I.

[37] Gaussian 16, Revision C.01, M.J. Frisch, G.W. Trucks, H.B. Schlegel, G.E. Scuseria, M.A. Robb, J.R. Cheeseman, G. Scalmani, V. Barone, G.A. Petersson, H. Nakatsuji, X. Li, M. Caricato, A.V. Marenich, J. Bloino, B.G. Janesko, R. Gomperts, B. Mennucci, H.P. Hratchian, J.V. Ortiz, A.F. Izmaylov, J.L. Sonnenberg, D. Williams-Young, F. Ding, F. Lipparini, F. Egidi, J. Goings, B. Peng, A. Petrone, T. Henderson, D. Ranasinghe, V.G. Zakrzewski, J. Gao, N. Rega, G. Zheng, W. Liang, M. Hada, M. Ehara, K. Toyota, R. Fukuda, J. Hasegawa, M. Ishida, T. Nakajima, Y. Honda, O. Kitao, H. Nakai, T. Vreven, K. Throssell, J.A. Montgomery, Jr., J.E. Peralta, F. Ogliaro, M.J. Bearpark, J.J. Heyd, E.N. Brothers, K.N. Kudin, V.N. Staroverov, T.A. Keith, R. Kobayashi, J. Normand, K. Raghavachari, A.P. Rendell, J.C. Burant, S.S. Iyengar, J. Tomasi, M. Cossi, J.M. Millam, M. Klene, C. Adamo, R. Cammi, J.W. Ochterski, R.L. Martin, K. Morokuma, O. Farkas, J.B. Foresman, D.J. Fox, Gaussian, Inc., Wallingford CT, 2019.

[38] J.J.P. Steward, Optimization of parameters for semiempirical methods VI: more modifications to the NDDO approximations and re-optimization of parameters, J. Mol. Model. 19 (2013) 1-32, https://doi.org/10.1007/s00894-012-1667-x.

[39] M.D. Hanwell, D.E. Curtis, D.C. Lonie, T. Vandermeersch, E. Zurek, G.R. Hutchison, Avogadro: an advanced semantic chemical editor, visualization, and analysis platform, J. Cheminf. 4 (2012) 17, https://doi.org/10.1186/1758-2946-4-17.

[40] W. Massa, Crystal Structure Determination, Springer-Verlag, Berlin Heidelberg 2000, https://doi.org/10.1007/978-3-662-04248-9.

[41] R.S. Rowland, R. Taylor, Intermolecular Nonbonded Contact Distances in Organic Crystal Structures: Comparison with Distances Expected from van der Waals Radii, J. Phys. Chem. 100 (1996) 7384-7391, https://doi.org/10.1021/jp953141+.

[42] Y. Yamamura, T. Murakoshi, M. Hishida, K. Saito, Examination of molecular packing in orthogonal smectic liquid crystal phases: a guide for molecular design of functional smectic phases, Phys. Chem. Chem. Phys. 19 (2017) 25518-25526, https://doi.org/10.1039/C7CP04744D.

[43] M.D. Ossowska-Chruściel, *Otrzymywanie i badania ciekłokrystalicznych tioestrów* [Eng. *Preparation and studies of liquid crystalline thioesters*], habilitation thesis, University of Siedlce, Siedlce 2008.

[44] J. Doucet, A.M. Levelut, M. Lambert, Polymorphism of the Mesomorphic Compound Terephthal-bis-Butylaniline (TBBA), Phys. Rev. Lett. 32 (1974) 301-303, https://doi.org/10.1103/PhysRevLett.32.301.

[45] J. Doucet, P. Keller, A.M. Levelut, P. Porquet, Evidence of two new ordered smectic phases in ferroelectric liquid crystals, J. Phys. France 39 (1978) 548-553, https://doi.org/10.1051/jphys:01978003905054800.

[46] M. Jasiurkowska, A. Budziak, J. Czub, S. Urban, Dielectric and X-ray Studies of Eleventh and Twelfth Members of Two Isothiocyanato Mesogenic Compounds, Acta Phys. Pol. A 110 (2006) 795-805, http://dx.doi.org/10.12693/APhysPolA.110.795.

[47] E. Dryzek, E. Juszyńska, R. Zaleski, B. Jasińska, M. Gorgol, M. Massalska-Arodź, Positron annihilation studies of 4-*n*-butyl-4'-isothiocyanato-1,1'-biphenyl, Phys. Rev. E 88 (2013) 022504, https://doi.org/10.1103/PhysRevE.88.022504.

[48] N. Bielejewska, E. Chrzumnicka, E. Mykowska, R. Przybylski, M. Szybowicz, Comparative Study of Orientational Order of Some Liquid Crystals from Various Homologous Series, Acta Phys. Pol. A 110 (2006) 777-793, http://dx.doi.org/10.12693/APhysPolA.110.777.

[49] N. Yadav, V. Swaminathan, V.P. Panov, R. Dhar, J.K. Vij, Elucidation of the de Vries behavior in terms of the orientational order parameter, apparent tilt angle, and field-induced tilt angle for smectic liquid crystals by polarized infrared spectroscopy, Phys. Rev. E 100 (2019) 052704, https://doi.org/10.1103/PhysRevE.100.052704.

[50] J. Budai, R. Pindak, S.C. Davey, J.W. Goodby, A structural investigation of the liquid crystal phases of 4-(2'-methylbutyl)phenyl 4'-n-octylbiphenyl-4-carboxylate, J. Phys. Lett. 45 (1984) 1053-1062, https://doi.org/10.1051/jphyslet:0198400450210105300.

[51] S. Lalik, O. Stefańczyk, D. Dardas, A. Deptuch, T. Yevchenko, S.-i. Ohkoshi, M. Marzec, Nanocomposites Based on Antiferroelectric Liquid Crystal (S)-MHPOBC Doping with Au Nanoparticles, Molecules 27 (2022) 3663, https://doi.org/10.3390/molecules27123663.

[52] B.M. Ocko, A.R. Kortan, R.J. Birgeneau, J.W. Goodby, A high resolution X-ray scattering study of the phases and phase transitions in N-(4-n-butyloxybenzylidene)-4-n-heptylaniline (40.7), J. Phys. 45 (1984) 113-128, https://doi.org/10.1051/jphys:01984004501011300.

[53] K. El Guermai, M. Ayadi, K. El Boussiri, Correlation Length in PBnA Liquid Crystal Family, Acta Phys. Pol. A 94 (1998) 779-784, http://dx.doi.org/10.12693/APhysPolA.94.779.

[54] K.Ch. Dey, P.Kr. Mandal, R. Dąbrowski, Effect of lateral fluorination in antiferroelectric and ferroelectric mesophases: synchrotron X-ray diffraction, dielectric spectroscopy and electro-optic study, J. Phys. Chem. Sol. 88 (2016) 14-23, https://doi.org/10.1016/j.jpcs.2015.08.016.

[55] K.Ch. Dey, P.Kr. Mandal, P. Kula, Effect of fluorinated achiral chain length on structural, dielectric and electro-optic properties of two terphenyl based antiferroelectric liquid crystals, J. Mol. Liq. 298 (2020) 112056, https://doi.org/10.1016/j.molliq.2019.112056.

**Structural investigation of the liquid crystalline phases of three homologues from the nOS5 series (n = 9, 10, 11) by X-ray diffraction**

Aleksandra Deptuch[1,*], Bartosz Sęk[2], Sebastian Lalik[3], Mirosława D. Ossowska-Chruściel[4], Janusz Chruściel[4], Monika Marzec[3]

[1] Institute of Nuclear Physics Polish Academy of Sciences, Radzikowskiego 152, PL-31342 Kraków, Poland
[2] Faculty of Physics and Applied Computer Science, AGH University of Kraków, Reymonta 19, PL-30059 Kraków, Poland
[3] Institute of Physics, Jagiellonian University, Łojasiewicza 11, PL-30348 Kraków, Poland
[4] Institute of Chemistry, Siedlce University of Natural Sciences and Humanities, 3 Maja 54, Siedlce PL-08110, Poland
* corresponding author, aleksandra.deptuch@ifj.edu.pl

# Supporting Information

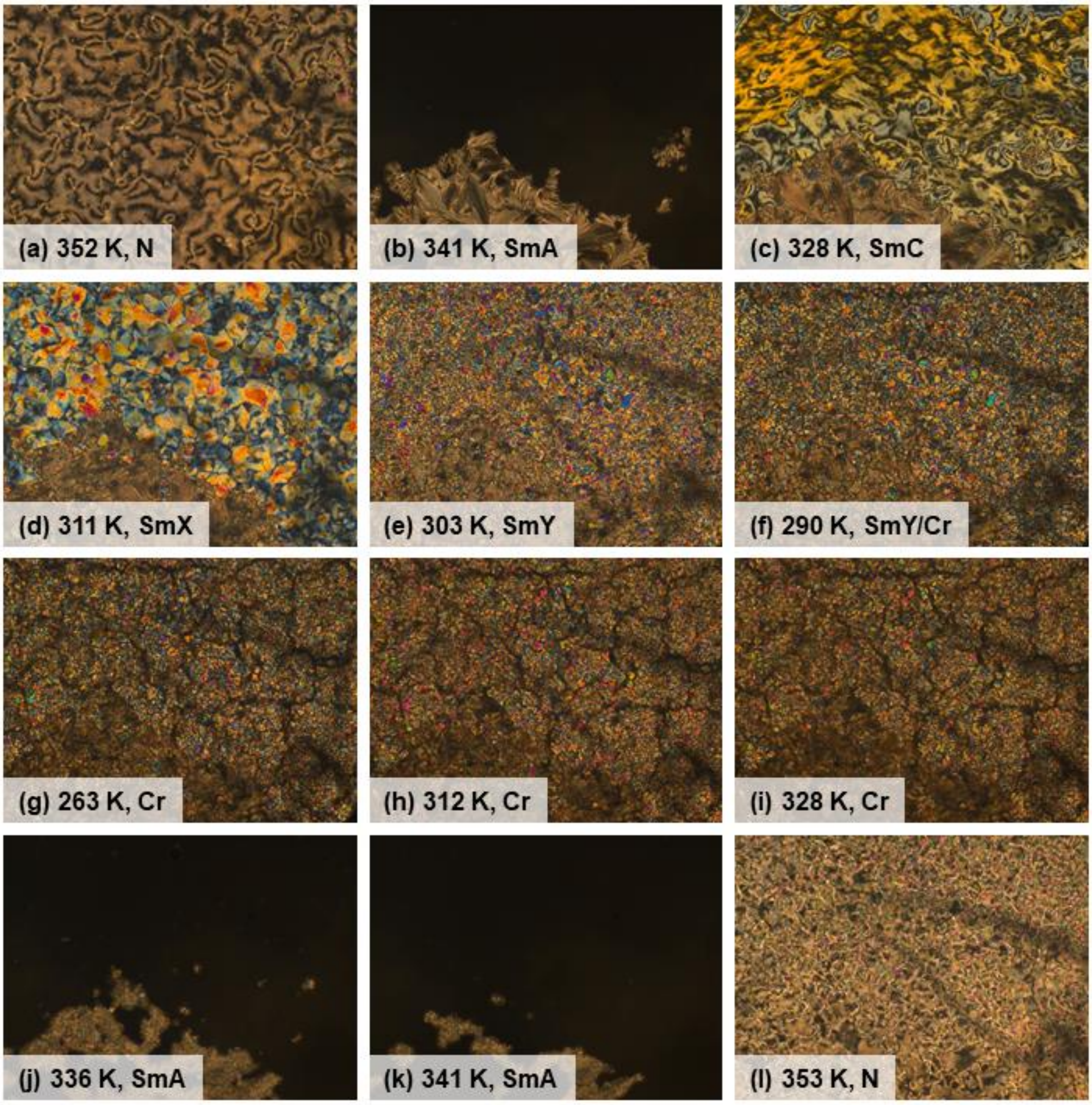

Figure S1. POM textures of 9OS5 collected during cooling (a-g) and heating (h-l) with the 5 K/min rate. Each photograph show an area of 1243 × 933 μm².

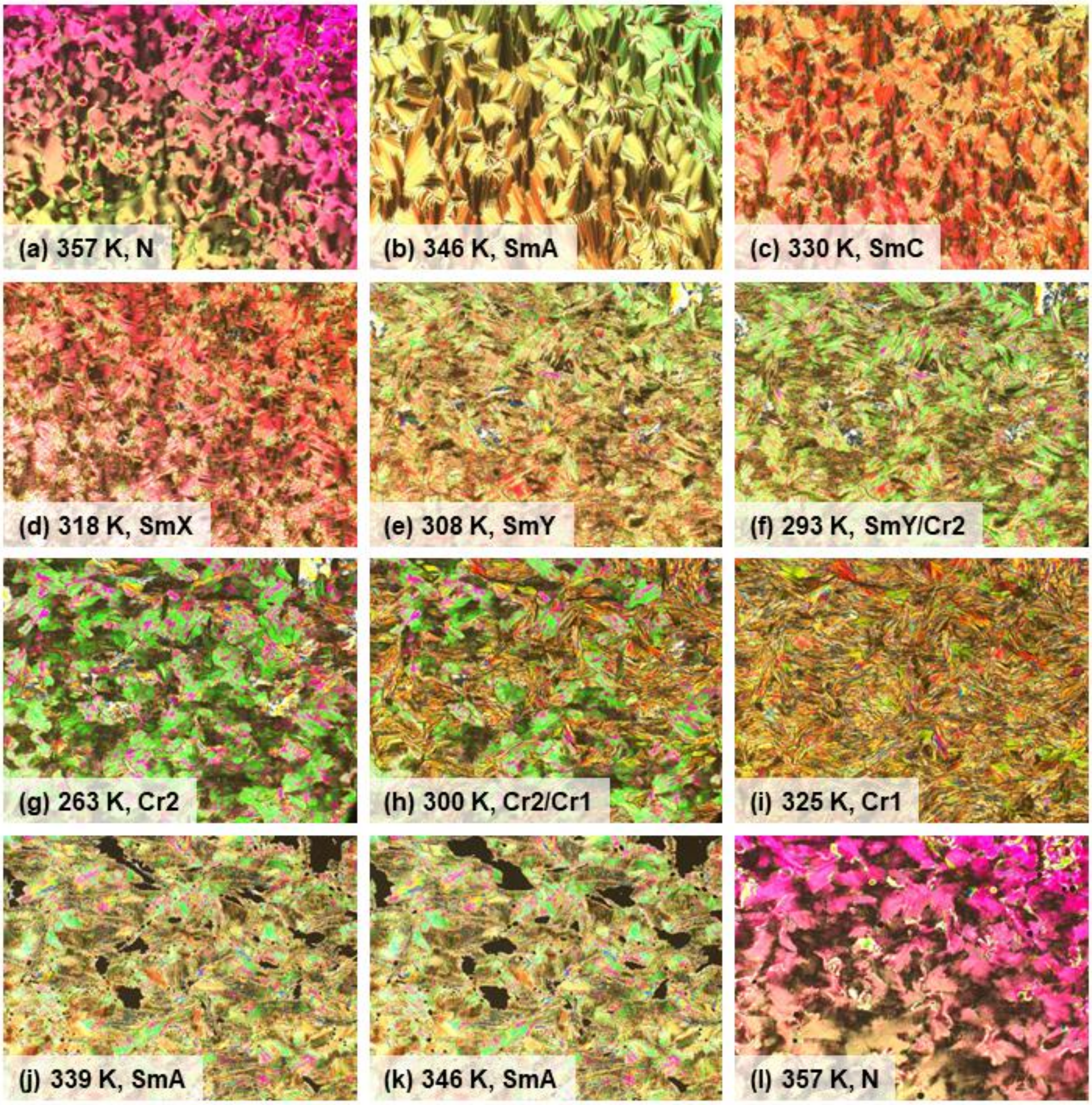

Figure S2. POM textures of 10OS5 collected during cooling (a-g) and heating (h-l) with the 5 K/min rate. Each photograph show an area of 1243 × 933 μm$^2$.

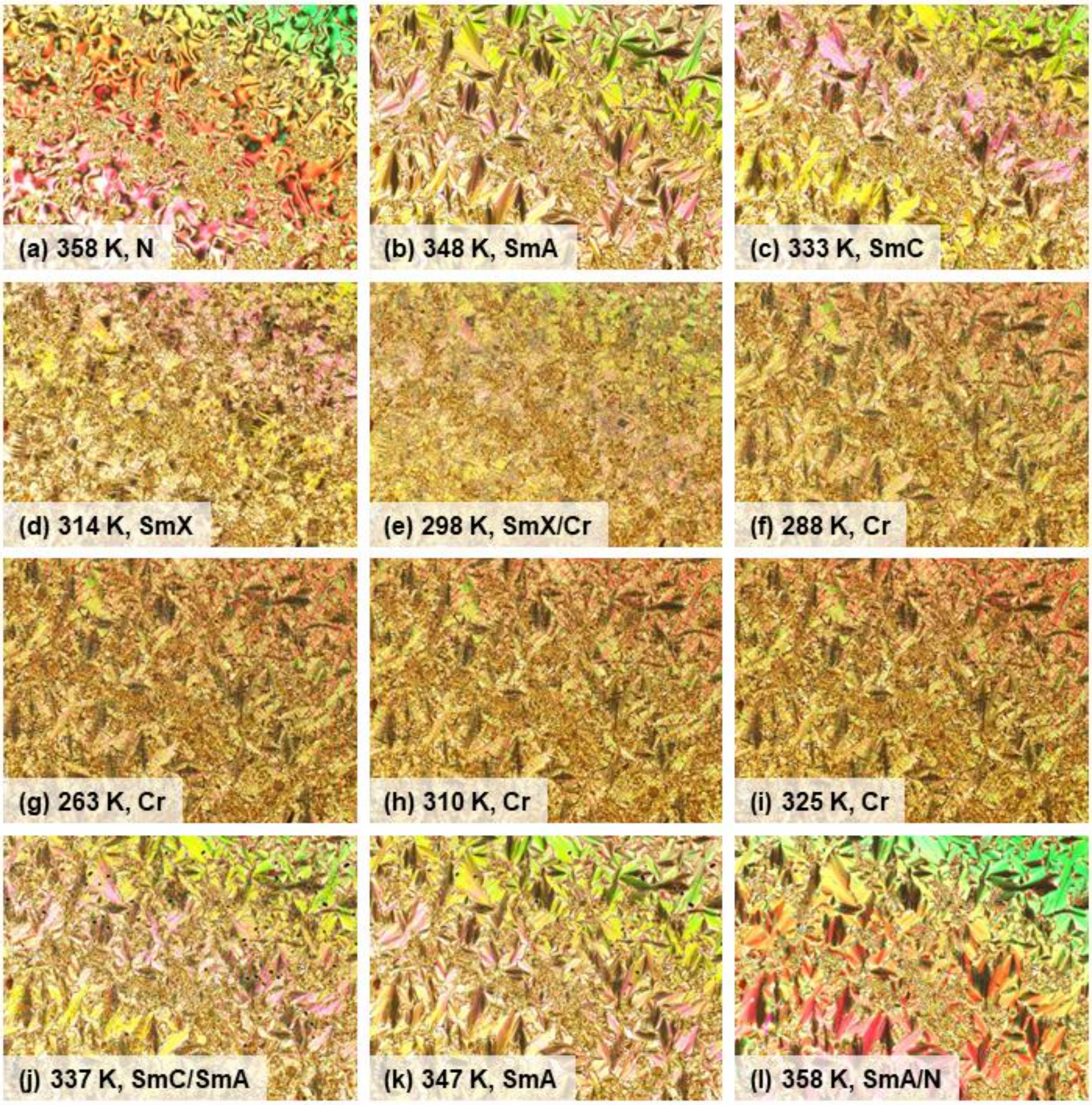

Figure S3. POM textures of 11OS5 collected during cooling (a-g) and heating (h-l) with the 5 K/min rate. Each photograph show an area of 1243 × 933 $\mu m^2$.

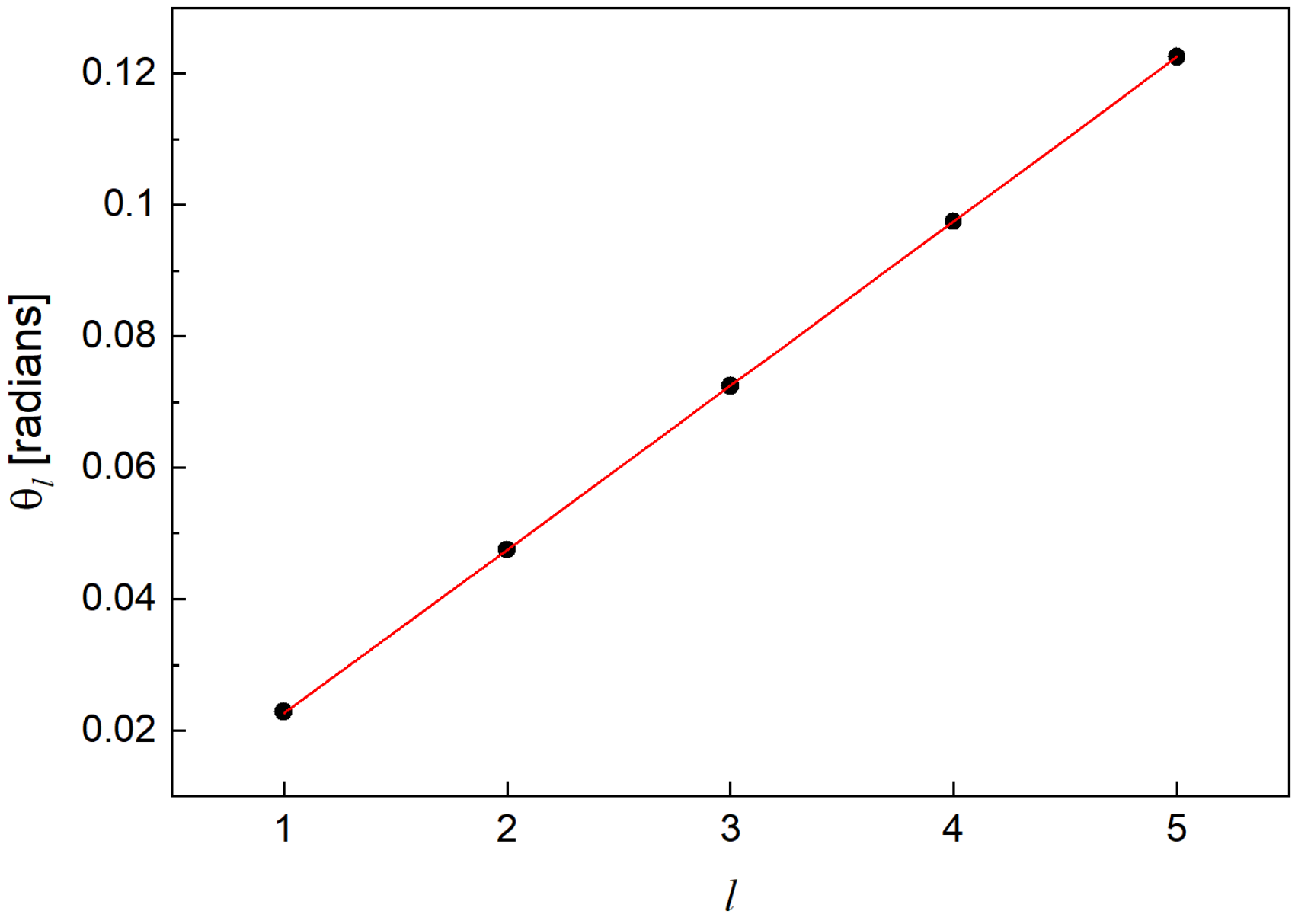

Figure S4. Determination of the layer spacing $d_{001}$ in the crystal phase of 10OS5 after slow cooling to 298 K from the positions of five diffraction peaks with the Miller indices from (001) to (005) by fitting the (2) formula from the main text.

### Calculation of the SmG unit cell

The formula for the interplanar distance $d_{hkl}$ in the monoclinic system, where $hkl$ are the Miller indices of the crystallographic plane, is as follows [I]:

$$\frac{1}{d_{hkl}^2} = \left(\frac{h}{a\sin\beta}\right)^2 + \left(\frac{k}{b}\right)^2 + \left(\frac{l}{c\sin\beta}\right)^2 - \frac{2hl\cos\beta}{ac\sin^2\beta}. \quad \text{(S1)}$$

The known distances, determined from the XRD patterns, are $d_{001}$, $d_{20\bar{2}}$, $d_{11\bar{1}}$, $d_{110}$. The resulting relationships are as follows:

$$c\sin\beta = d_{001}, \quad \text{(S2)}$$

$$\frac{1}{d_{110}^2} = \frac{1}{a^2\sin^2\beta} + \frac{1}{b^2}, \quad \text{(S3)}$$

$$\frac{1}{d_{11\bar{1}}^2} = \frac{1}{d_{110}^2} + \frac{1}{d_{001}^2} + \frac{2\cos\beta}{ac\sin^2\beta}, \quad \text{(S4)}$$

$$\frac{1}{d_{20\bar{2}}^2} = \frac{4}{a^2\sin^2\beta} + \frac{4}{d_{001}^2} + \frac{8\cos\beta}{ac\sin^2\beta}. \quad \text{(S5)}$$

The $\beta$ angle can be obtained as:

$$\cos\beta = a\sin\beta\, d_{001}\left(\frac{1}{d_{11\bar{1}}^2} - \frac{1}{d_{110}^2} - \frac{1}{d_{001}^2}\right), \quad \text{(S6)}$$

where:

$$a\sin\beta = \left(\frac{1}{4d_{20\bar{2}}^2} + \frac{1}{d_{110}^2} - \frac{1}{d_{11\bar{1}}^2}\right)^{-\frac{1}{2}}. \quad \text{(S7)}$$

After calculation of $\beta$, the $a$, $b$ and $c$ parameters are obtained from the (S7), (S3) and (S2) formulas, respectively.

## Calculation of the SmH unit cell

The distances necessary for calculations are $d_{001}$, $d_{200}$, $d_{110}$, $d_{11\bar{1}}$ or $d_{20\bar{2}}$. The knowledge of the $d_{200}$ distance enables use of a simpler relationship:

$$a \sin\beta = 2d_{200}, \tag{S8}$$

then the $\beta$ angle is determined from the $d_{11\bar{1}}$ or $d_{20\bar{2}}$ distance:

$$\cos\beta = 2d_{200}d_{001}\left(\frac{1}{d_{11\bar{1}}^2} - \frac{1}{d_{110}^2} - \frac{1}{d_{001}^2}\right), \tag{S9}$$

$$\cos\beta = \frac{d_{200}d_{001}}{4}\left(\frac{1}{d_{20\bar{2}}^2} - \frac{1}{d_{200}^2} - \frac{4}{d_{001}^2}\right), \tag{S10}$$

and the $a$, $b$, $c$ parameters are obtained from (S8), (S3) and (S2).

## Calculation of the SmJ and SmK unit cells

The necessary distances are $d_{001}$, $d_{020}$, $d_{110}$, $d_{11\bar{1}}$. The $b$ lattice constant can be directly obtained as:

$$b = 2d_{020}. \tag{S11}$$

The (S4) formula for the $c$ constant still holds, while the $a$ constant is related to the $\beta$ angle by the formula:

$$a \sin\beta = \frac{bd_{110}}{\sqrt{b^2 - d_{110}^2}}. \tag{S12}$$

The $\beta$ angle is determined using the $d_{11\bar{1}}$ distance:

$$\cos\beta = \frac{bd_{110}d_{001}}{2\sqrt{b^2 - d_{110}^2}}\left(\frac{1}{d_{11\bar{1}}^2} - \frac{1}{d_{110}^2} - \frac{1}{d_{001}^2}\right). \tag{S13}$$

After calculation of $\beta$, the $a$ and $c$ lattice constants can be eventually obtained using the (S12) and (S2) formulas.

[I] W. Massa, *Crystal Structure Determination*, Springer-Verlag, Berlin Heidelberg 2000, https://doi.org/10.1007/978-3-662-04248-9.